# Electron and phonon properties and gas storage in carbon honeycomb

Yan Gao[1], Yuanping Chen[1]*, Chengyong Zhong[1], Zhongwei Zhang[1],

Yuee Xie[1], and Shengbai Zhang[2]†

[1]*Department of Physics, Xiangtan University, Xiangtan, 411105, Hunan, China*

[2]*Department of Physics, Applied Physics, and Astronomy Rensselaer Polytechnic Institute, Troy, New York, 12180, USA*

## Abstract

A new kind of three-dimensional carbon allotropes, termed carbon honeycomb (CHC), has recently been synthesized [PRL 116, 055501 (2016)]. Based on the experimental results, a family of graphene networks are constructed, and their electronic and phonon properties are studied by various theoretical approaches. All networks are porous metal with two types of electron transport channels along the honeycomb axis and they are isolated from each other: one type of channels is originated from the orbital interactions of the carbon zigzag chains and is topologically protected, while the other type of channels is from the straight lines of the carbon atoms that link the zigzag chains and is topologically trivial. The velocity of the electrons can reach ~$10^6$ m/s. Phonon transport in these allotropes is strongly anisotropic, and the thermal conductivities can be very low when compared with graphite by at least a factor of 15. Our calculations further indicate that these porous carbon networks possess high storage capacity for gaseous atoms and molecules in agreement with experiment.

Corresponding authors: chenyp@xtu.edu.cn; zhangs9@rpi.edu.

# Introduction

Diamond and graphite are two well-known carbon allotropes. Their drastically different physical properties due to different atomic structures are a strong hint of the diverseness in the properties of carbon materials.[1-5] Over last several decades, a number of carbon allotropes have been found and synthesized, from zero dimensional (0D) fullerene to one dimensional (1D) carbon annotate to two dimensional (2D) graphene.[2, 6-9] The excellent properties of these carbon nanostructures place the study of carbon at the forefront of chemistry, materials science, and condensed matter physics.[10-16] In recent years, graphene and its derivative have emerged as the focus.[15, 17-21] Besides the 2D structures, 3D graphene networks have also been proposed,[22-30] which not only inherit the excellent properties of the graphene, but may offer other novel properties not seen in graphene. For example, the Mackay-Terrenes crystal, a kind of 3D graphene network, has been identified as a topological nodal-line semimetal.[24] Our recent studies indicate that a family of interpenetrated graphene network (IGN)[26,27] could be a Weyl-loop or a Weyl-surface semimetal, as well as a direct-gap semiconductor for optoelectronics when compressed into a carbon Kagome lattice (CKL).[28,29] A graphene network grown by chemical vapor deposition showed a very high electrical conductivity in line with the prediction of a high density of conducting electrons at the Fermi level.[22] Moreover, all the 3D networks have porous structures which could be ideal for energy storage, molecular sieves, and catalysis.[31-34]

Very recently, a new graphene network, termed carbon honeycomb (CHC), was synthesized,[35] where carbon was evaporated in vacuum from thin carbon rods heated by electric current to obtain carbon films. The analysis by transmission electron microscopy (TEM) and low temperature high energy electron diffraction showed that the films can be either periodic or random; all are porous networks built exclusively by $sp^2$-bonded carbon atoms. The experiment reveals that these carbon allotropes show high-level of physisorption of various gaseous molecules unattainable in other forms of carbon such as the fullerenes and nanotubes.[36,37] To date, however, the fundamental properties of the CHC, such as the electron and phonon properties, are still undetermined and the theoretical foundation for the observed gaseous absorption is also lacking.

In this paper, the electron and phonon properties and gas storage of the CHCs are studied by first-principles calculations. The results show that CHCs are rather stable carbon phases, whose cohesive energy, in the densest CHC-1 phase, is comparable to $C_{60}$ and CKL but higher than IGNs despite their remarkable structural similarity. As the density reduces with increased honeycomb cross section, the CHCs become more stable with energies close to diamond. All CHCs are metallic but with a mixed topological characteristics. Along the honeycomb axis, they inherit the super-high electron mobility from IGN and graphite. In the plane normal to the axis, on the other hand, there exists no electron transport channel. Phonon transport also exhibits a strong anisotropy, as phonon thermal conductivity normal to the honeycomb axis is only 1/3 of that along the axis. Even along the honeycomb axis, phonon thermal conductivity is exceptionally low. The large asymmetry between electron and phonon transport may be beneficial for thermoelectric applications.[38-40] For gas storage, on the other hand, the porous structures show a high capacity of inert gas and $CO_2$ adsorption, in agreement with experiment.[34]

## Model and Computational Methods

According to the suggestion in Ref. [35], a family of graphene networks can be constructed. All the structures consist of zigzag-edged graphene nanoribbons and every three adjacent nanoribbons are linked together by a line of $sp^2$ carbon atoms. Figures 1(a)-(b) show two examples, where carbon atoms of the nanoribbons (C2) are labeled blue while those of the linker (C1) are labeled purple. Looking from the top, the carbon atoms form honeycomb lattices of different sizes, which is the reason why they are termed CHCs. One can use CHC-$n$ to identify this family of carbon allotropes, where the index $n$ denotes the number of zigzag chains, i.e., the width of the nanoribbon. Therefore, Figs. 1(a)-(b) correspond to CHC-1 and CHC-2 with their primitive cells given in Figs. 1(c)-(d), respectively. The atomic structures for CHC-3 and CHC-4 are given in Figs. S1 in the supporting information (SI). It is noted that there are several structures in the literatures that are similar to the CHCs, such as IGN, CKL, and honeycomb carbon foam (HCCF) [26-30]. All of them are made of linked zigzag-edged graphene nanoribbons. The difference between the CHCs and the other structures is the lack of sp3 bonding in the former. In particular, the CHCs become HCCFs after the dimerization of those carbon atoms that connect the carbon ribbons[30].

Our first-principles calculations we rebased on the density functional theory within the PBE approximation for the exchange-correlation energy.[41-43] The core-valence interactions were described by the projector augmented-wave (PAW) potentials,[44] as implemented in the VASP code. Plane waves with a kinetic energy cutoff of 600 eV were used as the basis set. The calculations were carried out in periodic super cells. The atomic positions were optimized by using the conjugate gradient method, in which the energy convergence criterion between two consecutive steps was set at $10^{-5}$ eV. The maximum allowed force on the atoms is $10^{-2}$ eV/Å. The k-point meshes $7 \times 7 \times 13$ and $5 \times 5 \times 13$ were used for the Brillion Zone (BZ) integration of CHC-1 and CHC-2, respectively.[45] When calculating the adsorption of gaseous atoms/molecules, the Grimme-D3 correction was included in the calculation to account for van der Waals interactions.[46,47] Phonon spectra were obtained by using the Phonopy package.[48]

To calculate the phonon thermal conductivity, classical molecular dynamics (MD)[49] simulations were carried out by using the LAMMPS package[50] with the Adaptive Intermolecular Reactive Empirical Bond Order (AIREBO) potential,[51] which is proven to be effective for Van der Waals interactions and has reasonable accuracy for thermal properties of carbon allotropes[52,53]. In the simulations, the initial structures are relaxed in the NPT ensemble with a time step of 0.35 fs. After the relaxation, we calculated the heat flux under the NVE ensemble. By integrating the autocorrelation function of the heat flux, the thermal conductivity in different direction are extracted by using the Green-Kubo formula.[46]

## Results and Discussions

The structural properties for CHC-$n$ ($n$ = 1-4) are shown in Table 1. For comparison, the corresponding properties of HCCF, IGN, CKL, graphite and diamond are also show on. The CHCs have two distinct space groups: if $n$ is odd, it belongs to P6$_3$/mmc; if $n$ is even, it belongs to P6/mmm. The bond lengths of the CHCs vary in the range of 1.41 ~ 1.49Å, which can be compared to those of diamond(1.54 Å) and graphite (1.42 Å). The longer bonds are the bonds connecting blue and purple atoms. Hence, in spite of the three-fold coordination, some of the CHC's bond lengths can be closer to that of sp$^3$ bonds. The open porous structures result in small carbon densities, much smaller than graphite. However, all the CHCs have much larger bulk moduli than graphite. The calculated CHCs' cohesive energies $E_{coh}$ are large, being 0.4 eV smaller than diamond for CHC-1, but the difference decreases substantially when the honeycomb cross section increases. They are also noticeably smaller than IGNs' and HCCFs' (and their derivatives) but are comparable to CKLs' as the carbon densities are approximately equal.

Figure 2 shows the electronic properties of CHC-1, which appear complex with multiple bands cross the Fermi level, as revealed by the orbital-projected band structure in Fig. 2(a). The partial density of states (PDOS) profiles in Fig. 2(b) suggest that states near the Fermi level are mainly $p_z$ orbitals on the C1 atoms, corresponding to the two red bands, but $p_x$/$p_y$ orbitals on the C2 atoms, corresponding to the green bands. As such, there exist two distinct transport channels along the honeycomb axis: Fig. 2(c) shows the charge contour of the $F_1$ state in Fig. 2(a), corresponding to the C1 atoms; Fig. 2(d) shows that of the $F_2$ state corresponding to the C2 atoms on the zigzag chains. Although the two channels are adjacent to each other, they have practically no interactions, because spatially the C1 $p_z$ orbitals are normal to the C2 $p_x$/$p_y$ orbitals. Hence, the two conducting channels are isolated from each other. Regardless, electrons moving in either channel have very high speeds $\sim 10^6$ m/s, which is analogous to the Dirac electrons in graphene. The electronic structure of CHC-1 is also highly anisotropic as evidenced by Fig. 2(a): around the Fermi level, bands along the A − H and L − H directions are nearly flat. Hence, in the planes normal to the honeycomb axis, electrons move rather slowly.

According to the orbital characteristics of CHC-1, a tight-binding model is constructed to describe its electron properties,

$$H = \sum_{i,j} t_{1\alpha} a_i^\dagger a_j + \sum_{l,m} t_{2\beta} b_l^\dagger b_m, \ (\alpha, \beta = 1, 2) \qquad (1)$$

where $a_i^\dagger/b_l^\dagger$ and $a_j/b_m$ represent the creation and annihilation operators, $t_{1\alpha}$ ($\alpha = 1, 2$) are the hopping parameters between C1 atoms, and $t_{2\beta}$ ($\beta = 1, 2$) are those between C2 atoms, respectively. The hopping parameters are given in Fig. 3(a). We have omitted interactions between C1 and C2 atoms in Eq. (1). Figure 3(b) shows the tight-binding band structure, which agrees rather well with first-principles results in Fig. 2(a). Therefore, the model captures the essential physics of first-principles results, despite its simplicity.

Figure 4 shows the orbital-projected band structure and PDOS for CHC-2. Both of them are rather similar to CHC-1, which suggests that carrier transport properties in CHCs should all be similar, namely, highly anisotropic and having two distinct transport channels along the honeycomb axis. The main difference is that the two red bands originated from the $p_z$ orbitals of C1 atoms get closer in energy. This is because the distance between the C1 lines gets wider, which decreases the hopping energy $t_{12}$ in Eq. (1) and thus reduces the effective interaction between the two bands. Eventually, the two bands become nearly degenerate when $n$ is sufficiently large (as shown in Fig. S2 in SI).

As mentioned above, the CHCs possess structures that resemble the graphene networks proposed in Refs. [26-30], although their electronic properties on the appearance can be markedly different. The IGNs in Refs. [26, 27] are pure Weyl-like semimetals and may be viewed as also being made of the C2 zigzag-edged graphene nanoribbons for the CHCs. The band structures in Figs. 2(a) and 4(a) reveal that the green bands also have Weyl-like nodal points along $\Gamma - A$, as a reminiscence of the IGNs. Hence, the high mobility of the isolated C2 conduction channel in CHCs can be non-trivial and topologically protected. While the isolated C1 channel also shows a high conductivity, its band structure near the Fermi level is trivial. Hence, the CHCs represent an interesting mixture of two topologically different conduction channels (isolated from each other). Another interesting analogy between CHCs and the structures in Refs. [28,29], namely, the CKLs, obtained by a uniaxial compression of the IGNs, is the common existence of orbital frustration due to their triangular lattices. In the case of CHCs, it is the zigzag-edged nanoribbons that form the triangular lattice, leaving hidden orbital frustrations in their electronic structure, as can be seen by a wave function analysis in Fig. 2(e).[28]

Figures 5(a)-(b) show the calculated phonon dispersions of CHC-1 and CHC-2. From the strong anisotropy in the phonon dispersion, one can expect that phonon transport should also show a strong anisotropy, as can been seen in Fig. 5(c). Along the $\Gamma - A$ path, the acoustic braches and most of the optical branches show relatively large slopes. Therefore, phonons can transport along the honeycomb axis with a relatively high mobility – a property that is also inherited from graphene. However, in the directions normal to the honeycomb axis, group velocities of phonons are considerably smaller especially for optical phonons, as a result of the small

dispersions along $\Gamma - K$ and $\Gamma - M$. For CHC-1, the calculated thermal conductivity is 260 ± 5 W/mK along the honeycomb (z-) axis but only 80 ± 5 W/mK along any of the directions inside the xy plane. The thermal conductivities for CHC-2, on the other hand, are lower: 88 ± 3 W/mK along the z axis and 32 ± 3 W/mK in the *xy* plane. These values are considerably smaller than that of graphene (about 4,000 W/mK) by a factor of 15 to 125. It suggests that the formation of porous structures could be an effective way to reduce thermal conductivity without jeopardizing electrical transport.

Reference [35] suggests that CHCs could be good absorbers for $CO_2$ and other inert gas molecules such as Kr and Xe, which are unattainable under the same condition in other carbon forms such as fullerites and nanotubes.[25] To address this issue, we have considered gas adsorption in various CHCs. The adsorption energy is defined as $E_n = E_{tot} - E_{CHC} - E_{gas}$, where $E_{tot}$, $E_{CHC}$ and $E_{gas}$ are the total energies after and before adsorption, and that of the *n* atoms (molecules) in the gas phase, respectively. From $E_n$ which is the total adsorption energy, we further define the consecutive adsorption energy $\delta E_C = E_n - E_{n-1}$. Using the $1 \times 1 \times 1$ primitive cell of CHC-2, we found only He can be adsorbed exothermically with up to four atoms per primitive cell. However, the average adsorption energies $\delta E_A = E_n/n = -(2-3)$ meV/He are too small, due to the too-packed spacing along *z* (see Fig. S3 in SI). At half of the gas loading level, e.g., in the $1 \times 1 \times 2$ supercell, on the other hand, all the inert gases and $CO_2$ can be exothermically adsorbed with substantially lowered energies.

Figures 6 shows the adsorption sites for single atom/molecule, whereas Fig. S4 in SI shows the sites for multiple atoms/molecules. Table 2 summarizes the results for $\delta E_C$ and the maximum adsorption density, $\rho_{max}$. We see that the loading densities are 4, 3, 1, 1, 1, and 1, for He, Ne, Ar, Kr, Xe, and $CO_2$, respectively. At the single-atom loading level, $\delta E_C = -(46-81)$ meV/atom, which is substantial larger in magnitude than the experimental temperatures of 57 K ($k_B T = 5$ meV) for Kr and ~120 K ($k_B T = 11$ meV) for He and Xe. The energy for $CO_2$ of $-10$ meV/$CO_2$ is, however, substantially smaller, but its magnitude is still large rthan the experimental temperature of 80-85 K ($k_B T = 7$ meV). Reference [35] reported that He adsorption can reach 16 wt% in CHC-4. Our results in Table S1 in SI suggest that at $\rho = 17$ wt%, $\delta E_C = -16$ meV/He, whose magnitude is again noticeably larger than $k_B T = 11$ meV.

The maximum theoretical loading density is $\rho_{max} = 19.6$ wt% at which $\delta E_C$ remains to be negative, although at a smaller value of $-7$ meV/He. The significant energy lowering here is due to CHC: as a check, we removed the CHC in the case of four He in Table 2 and found $E_{n=4} = -6$ meV so $\delta E_A = -1.5$ meV/He, which means, in our calculation, the attraction between He atoms is negligibly small. In addition, we included the D3 correction in our calculations but the results show that its effect is less than 1 meV per adsorbed atom.

## Conclusion

We have studied the electron and phonon transport properties of the recently discovered CHCs. Both properties exhibit strong anisotropy reflecting their geometric anisotropy. There are two types of distinct electron transport channels along the honeycomb axis; both having a high carrier mobility. One type of the channels is in fact topologically protected similar to the IGNs proposed earlier, while the other type is topologically trivial. For phonon transport, thermal conductivity along the honeycomb axis is very low comparing to graphite. This is because the porous geometries result in a larger cross-sectional area for phonon scattering. The thermal conductivity normal to the axis is even lower by a factor 1/3. We have also studied gas storage capacity. The results show a high gas loading densities of CHCs for He, Ne, Ar, Kr, Xe, and $CO_2$, in agreement with experiment.

## Acknowledgments

This work was supported by the National Natural Science Foundation of China (Nos.51376005and 11474243). SZ acknowledges the support by US DOE under Grant No. DE-SC0002623.

**Figure captions**

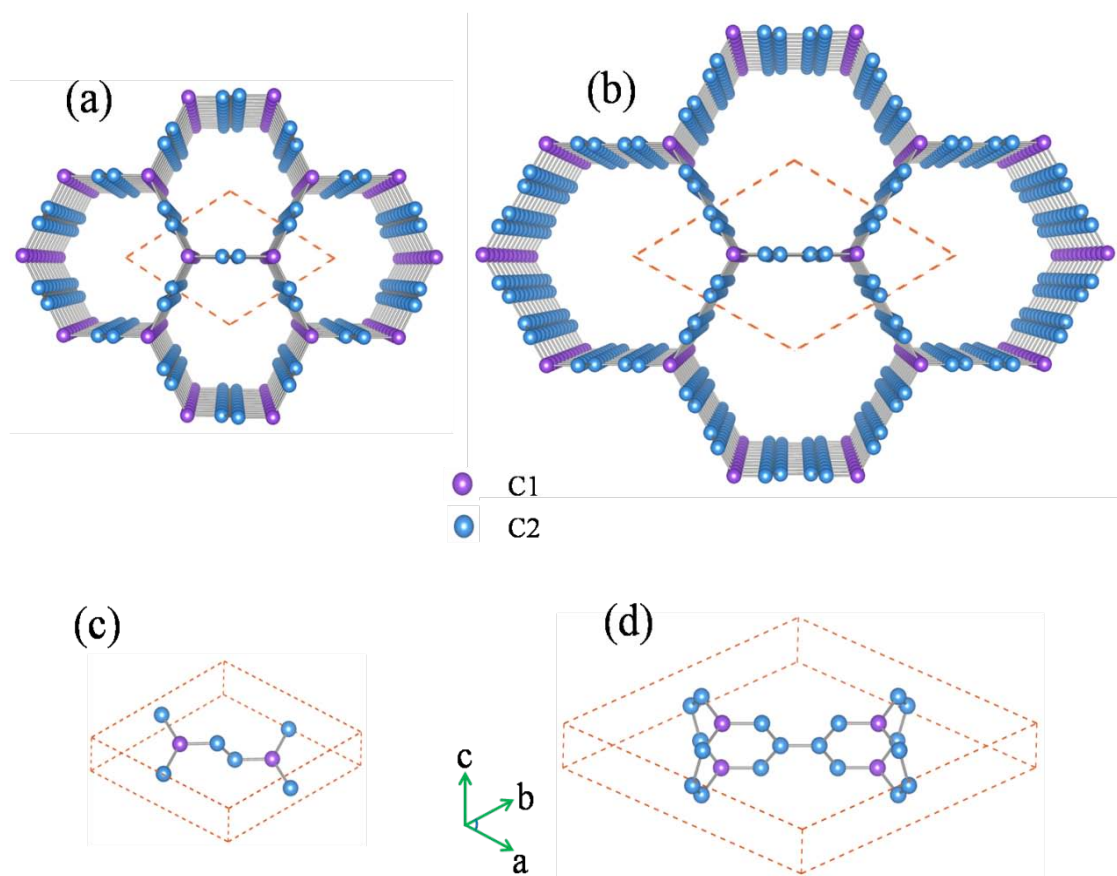

Figure 1 Atomic structure of (a) CHC-1 and (b) CHC-2, consisting of zigzag-edged graphene nanoribbons (blue C2 atoms) and linking lines (purple C1 atoms). The dashed lines show the top views of primitive cells, while the side views of the primitive cells are shown in (c) and (d), respectively.

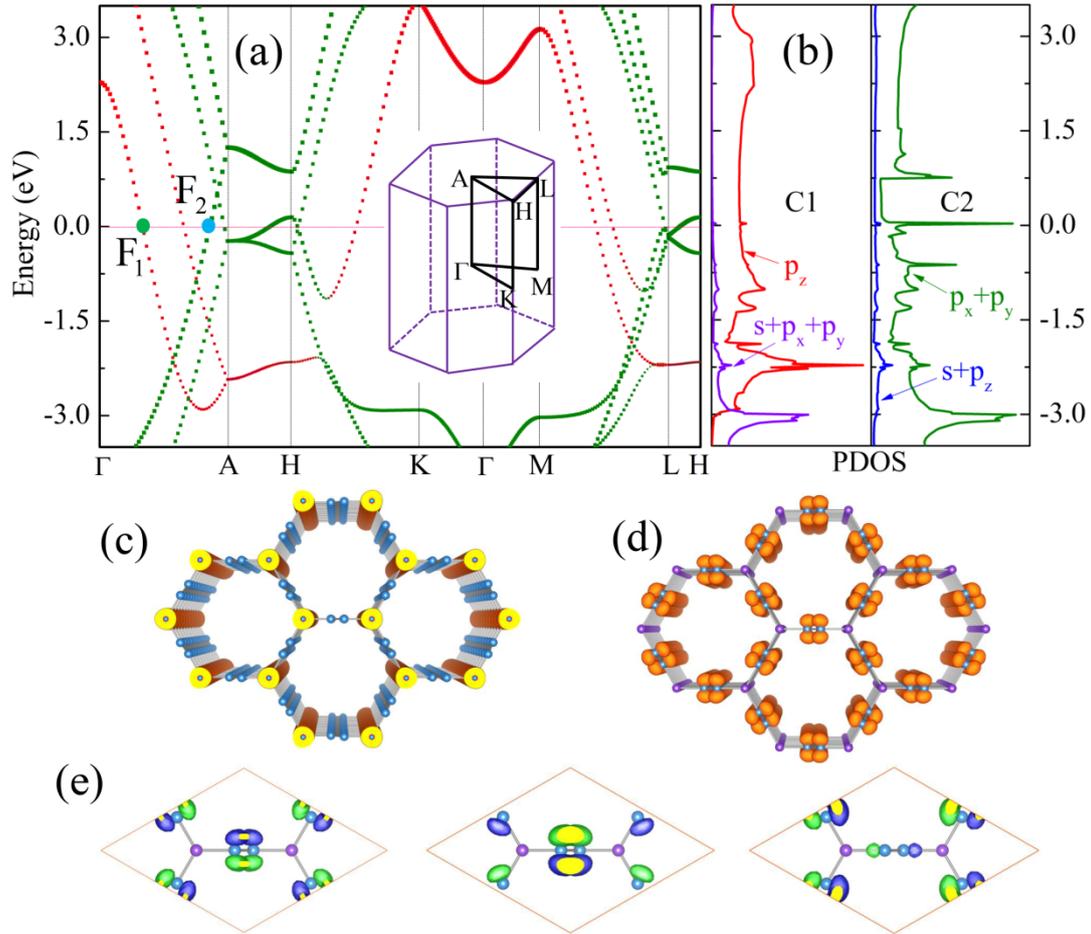

Figure 2 (a) Orbital projection band structure of CHC-1, where the red and green bands correspond to $p_z$ orbitals of C1 atoms and $p_x/p_y$ orbitals of C2 atoms. Inset: BZ of CHC-1. (b) PDOS of C1 and C2 atoms in CHC-1. (c-d) Charge densities of states at F1 and F2 points in (a). (e) Three orbital frustration states in CHC-1, which are states in the green bands (a) at the Γ point with energies -5.161 and -3.856 eV. The first one is an anti-bonding state at -5.161 eV, while the last two states are degenerated states at -5.161 eV.

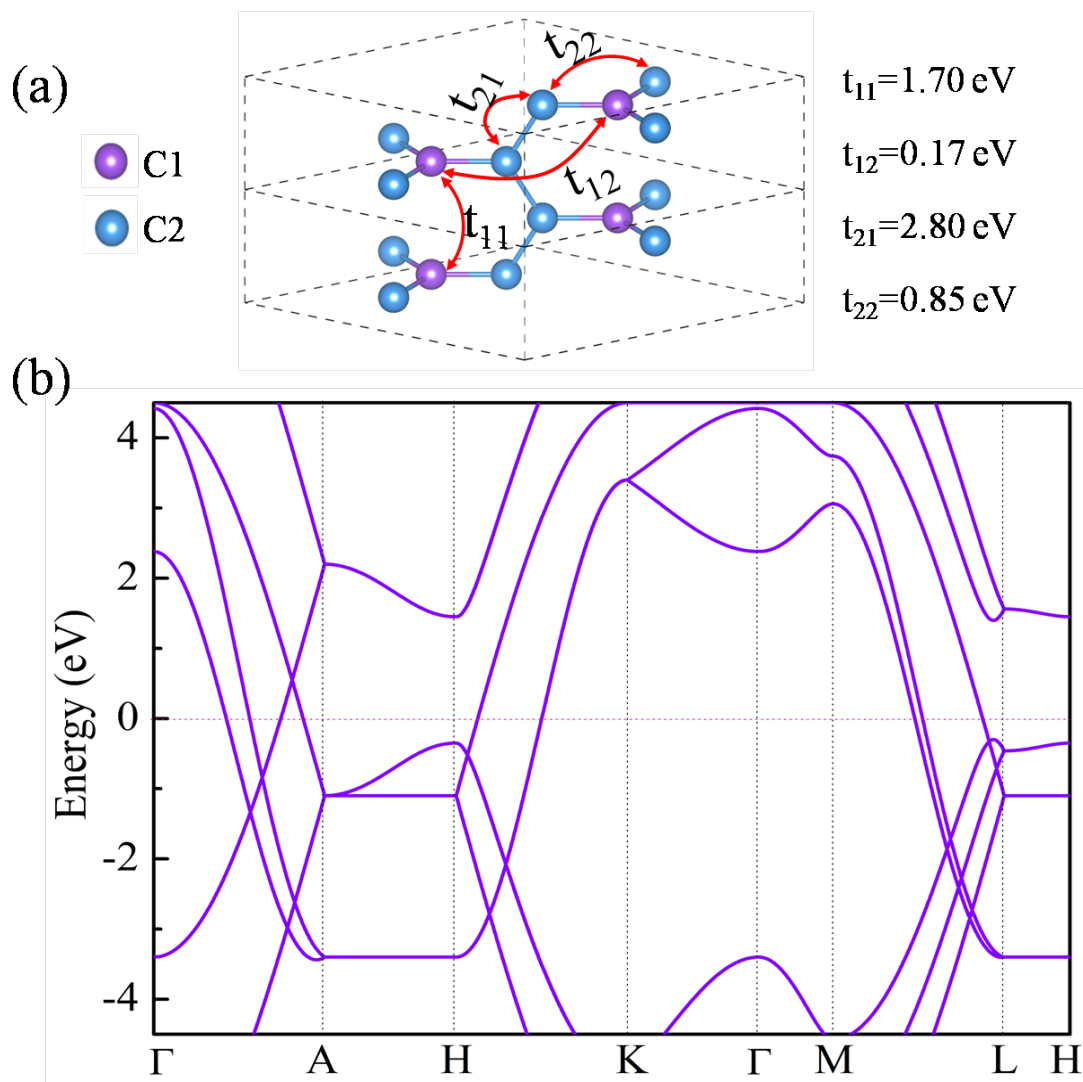

Figure 3 (a) Tight-binding parameters in the primitive cell of CHC-1. (b) Tight-binding band structure of CHC-1.

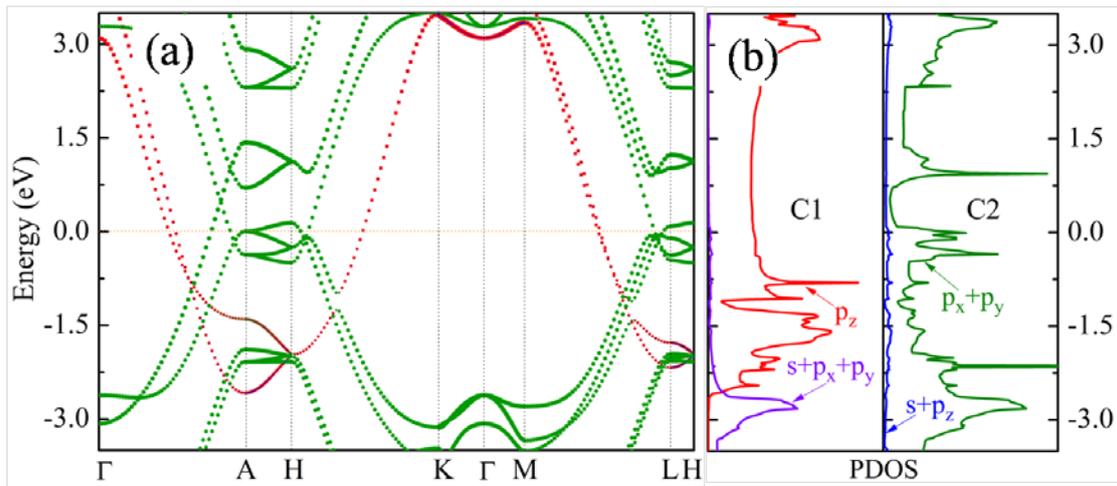

Figure 4 Orbital projection band structure (a) and PDOS (b) of CHC-2.

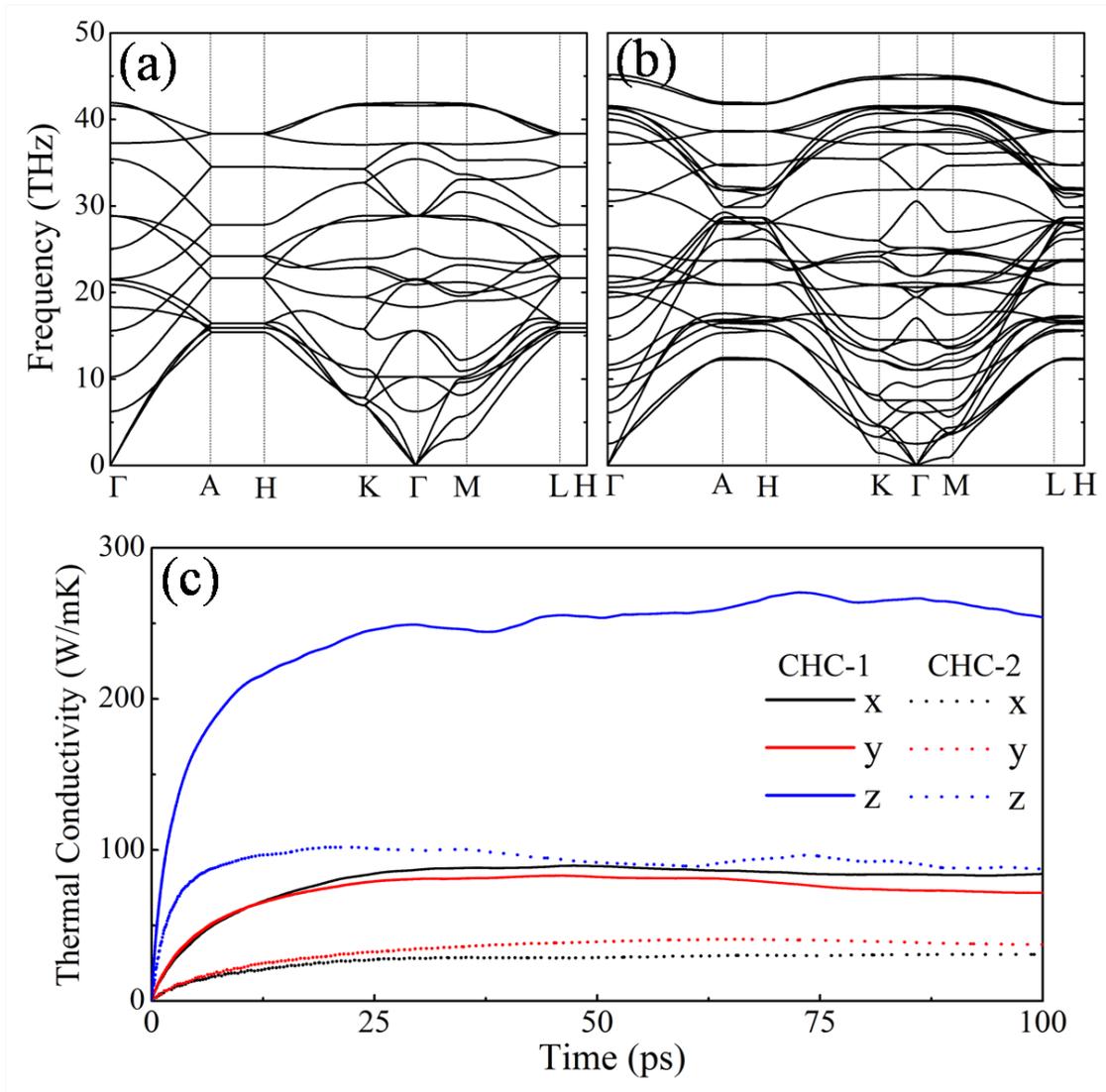

Figure 5 Phonon dispersions of (a) CHC-1 and (b) CHC-2. (c) Time dependence of phonon thermal conductivities of CHC-1 and CHC-2 along axes x, y and z.

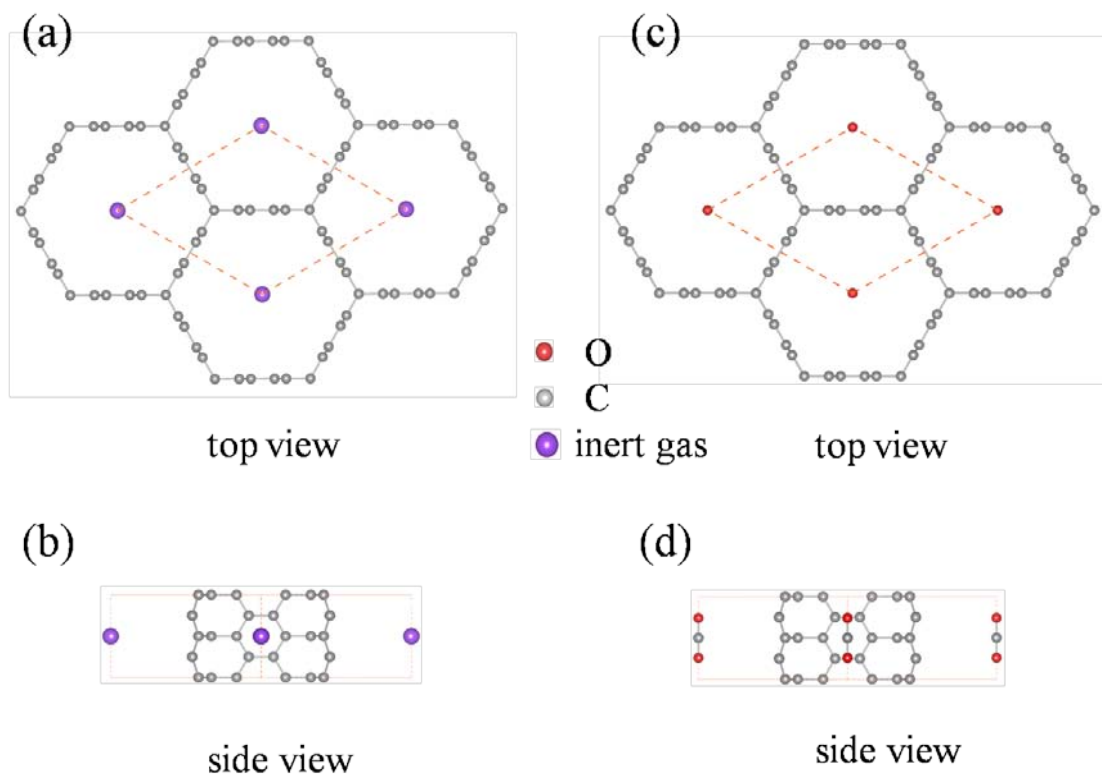

Figure 6 Top view (a) and side view (b) of CHC-2 as one inert gas atom per $1\times1\times2$ super cell adsorbs. (c) and (d) are for $CO_2$ adsorption.

Table 1 Space group, lattice parameters (Å), density (g/cm$^3$), bond lengths (Å), bulk moduli (GPa), and cohesive energy $E_{coh}$ (eV/C) for CHC-$n$ ($n$ = 1~4), HCCF-1, CKL-1, IGN-1, diamond, and graphite. HCCF-1, CKL-1, IGN-1, consisting of $n$ = 1 nanoribbons, are the simplest structures of their families

| Structure | Space group | Lattice parameters | | Density | Bond lengths | Bulk Moduli | $E_{coh}$ |
|---|---|---|---|---|---|---|---|
| | | a = b | c | | | | |
| CHC-1 | P6$_3$/mmc | 6.39 | 2.42 | 1.87 | 1.41, 1.49 | 220.8 | -7.37 |
| CHC-2 | P6/mmm | 10.12 | 2.43 | 1.29 | 1.41, 1.45, 1.49 | 153.3 | -7.55 |
| CHC-3 | P6$_3$/mmc | 13.84 | 2.44 | 0.98 | 1.41~1.49 | 117.5 | -7.65 |
| CHC-4 | P6/mmm | 17.54 | 2.45 | 0.79 | 1.41~1.49 | 95.1 | -7.71 |
| HCCF-1 | P-3m1 | 6.35 | 4.83 | 1.89 | 1.38~1.64 | 225.0 | -7.57 |
| CKL-1 | P6$_3$/mmc | 4.46 | 2.53 | 2.75 | 1.50, 1.53 | 322.1 | -7.44 |
| IGN-1 | CMCM | 4.33 | 2.47 | 2.59 | 1.41, 1.53 | 311.4 | -7.62 |
| Diamond | Fd$\bar{3}$m | 3.56 | 3.56 | 3.55 | 1.54 | 431.3 | -7.77 |
| Graphite | P6$_3$/mmc | 2.46 | 6.80 | 2.24 | 1.42 | 36.4 | -7.90 |

Table 2. Consecutive adsorption energy $\delta E_C$ for inert gas atoms and $CO_2$ loading of CHC-2. All energies are given in meV/adsorbent. $\rho_{max}$ is also given in atomic weight percent.

| #adsorbent | He | Ne | Ar | Kr | Xe | $CO_2$ |
|---|---|---|---|---|---|---|
| 1 | -46 | -48 | -58 | -67 | -81 | -10 |
| 2 | -13 | -19 | +25 | > 0 | > 0 | > 0 |
| 3 | -18 | -16 | – | – | – | – |
| 4 | -5 | +27 | – | – | – | – |
| 5 | +26 | – | – | – | – | – |
| $\rho_{max}$ (wt%) | 4.55 | 15.27 | 10.64 | 19.96 | 28.10 | 11.58 |